\documentclass[runningheads]{svmult}

\usepackage{makeidx}   % allows index generation
\usepackage{graphicx}  % standard LaTeX graphics tool
\usepackage{subeqnar}  % subnumbers individual equations
\usepackage{multicol}  % used for the two-column index
\usepackage{physprbb}  % modified textarea for proceedings,
\makeindex             % used for the subject index
\newcommand{\Ha}{\mbox{H$\alpha$}}
\newcommand{\OII}{\mbox{[OII]}}
\newcommand{\micron}{$\mu$m} 
\newcommand{\mifs}{\texttt{MIFS}}
\newcommand{\sauron}{\texttt{SAURON}}
\newcommand{\oasis}{\texttt{OASIS}}
\newcommand{\tiger}{\texttt{TIGER}}
\newcommand{\vimos}{\texttt{VIMOS}}

\newcommand{\gmos}{\texttt{GMOS}}
\newcommand{\threed}{\texttt{3D}}
\newcommand{\sinfoni}{\texttt{SINFONI}}
\newcommand{\stis}{\texttt{STIS}}
\def\arcsec{\hbox{$^{\prime\prime}$}}

\def\farcs{\hbox{$.\!\!^{\prime\prime}$}}
\def\ergs{\hbox{erg~s$^{-1}~\!$cm$^{-2}$}}
\def\Msun{\hbox{M$_\odot$}}

\begin{document}
\title*{A Mega Integral Field Spectrograph for the VLT}
\toctitle{A Mega Integral Field Spectrograph for the VLT}
%\protect\newline in the Particle Deflection Plane}
% allows explicit linebreak for the table of content
%
%
\titlerunning{MIFS: A Mega Integral Field Spectrograph for the VLT}
% allows abbreviation of title, if the full title is too long
% to fit in the running head
%
\author{R.\ Bacon\inst{1}
\and G.\ Adam\inst{1}
\and S.\ Cabrit\inst{3}
\and F.\ Combes\inst{3}
\and R.L.\ Davies\inst{2}
\and E.\ Emsellem\inst{1}
\and\\ P.\ Ferruit\inst{1}
\and M.\ Franx\inst{5}
\and G.\ Gilmore\inst{7}
\and B.\ Guiderdoni\inst{4}
\and O.\ Lef\`evre\inst{6}
\and S.\ Morris\inst{2}
\and\\ E.\ P\'econtal\inst{1}
\and E.\ Prieto\inst{6}
\and R.\ Sharples\inst{2}
\and P.\ van der Werf\inst{5}
\and P.T.\ de Zeeuw\inst{5}}
\authorrunning{Bacon et al.}
% if there are more than two authors,
% please abbreviate author list for running head
%
%
\institute{CRAL, Observatoire de Lyon, 69230 Saint-Genis-Laval Cedex, France
\and Physics Department, University of Durham, Durham DH13LE, UK
\and Observatoire de Paris, 61 avenue de l'Observatoire, 75014 Paris, France
\and Institut d'Astrophysique de Paris, 98bis bd Arago, 75014 Paris, France
\and Sterrewacht Leiden, Postbus 9513, 2333 CA Leiden, the Netherlands
\and Laboratoire d'Astrophysique de Marseille, 13376 Marseille Cedex 12, France
\and Institute of Astronomy, Madingley road, Cambridge CB3 0HA, UK
}

\maketitle              % typesets the title of the contribution

\begin{abstract}
We describe \mifs, a second generation integral-field spectrograph for
the VLT, operating in the visible wavelength range. It combines a
$1'\!\times \!1'$ field of view with the improved spatial resolution
provided by multi-conjugate adaptive optics and covers a large
simultaneous spectral range (0.6--1.0~\micron). A separate mode
exploits the highest spatial resolution provided by adaptive
optics. With this unique combination of capabilities, \mifs\ has a
wide domain of application and a large discovery potential.

The \mifs\ low-spatial resolution mode (sampled at 0\farcs2) combined
with the initial MCAO capabilities planned for the VLT will provide
ultra-deep fields with a limiting magnitude for spectroscopy of $R\sim
28$. \mifs\ will improve the present day detection limit of Ly$\alpha$
emitters by a factor of 100, and will detect low-mass star-forming
galaxies to $z \sim 7$. The \mifs\ high-spatial resolution mode
($3''\times 3''$ field sampled at 0\farcs01) is optimized for the next
step in (MC)AO. It will probe, e.g., the relationship between supermassive
central black holes and their host galaxy and the physics of winds
from accretion disks in young stellar objects at unprecedented spatial
resolution.

\mifs\ will extend Europe's lead in integral-field spectroscopy. It
capitalizes on new developments in adaptive optics, and is a key step
towards instrumentation for OWL.\looseness=-2
\end{abstract}

\section{Introduction}
\label{sec:intro}
The Hubble Deep Fields (HDFs)---the deepest broad band images obtained
to date---have revealed the distribution in redshift and morphology of
high redshift galaxies, and have generated an impressive number of
follow-up studies at all wavelengths, so that nearly complete
multi-wavelength imaging is available for distant galaxies. However,
many important physical quantities (gas/star content, stellar
population mix, kinematics) cannot be derived from broad-band imaging
but require spectroscopic measurements.  A significant investment of
observing time on large ground-based telescopes, in particular Keck,
has allowed completion of spectroscopic observations of the HDF north
galaxies to $R\sim 24$, most of which are at $z\!<\!1$.  To increase
the redshift limit, one needs to achieve a better detection limit,
higher spatial resolution, low sky background, and an ultra-stable and
well-calibrated instrument. These requirements are one of the central
drivers for NGST, to be launched in 2009.

We show here that ground-based astronomy can benefit from new
technological developments such as multi-conjugate adaptive optics
(MCAO), high-order deformable mirrors, second-generation image
slicers, and panoramic integral-field spectrographs (IFSs) to
challenge the capabilities of NGST well before 2009. We propose an
instrument for the VLT, \mifs, which will not only provide the
critical spectral information that is currently missing in our
understanding of high-$z$ galaxies, but will also allow major progress
in many other areas of astronomy.

\section{Design considerations}
\label{sec:design}
We first summarize recent technical developments which influence our
design for \mifs, then discuss the scientific tradeoff we conducted to
set its specifications, and conclude with a performance estimate.

\subsection{Developments in adaptive optics}
\label{sec:ao}
MCAO is being developed to overcome the problem of isoplanetic angle,
which affects classical AO. First generation MCAO uses three or five
reference stars to sense atmospheric turbulence in three dimensions,
and two deformable mirrors conjugated with the main layers of
turbulence. Compared to the classical (single layer) AO, MCAO
dramatically improves the corrected field of view. The ESO AO group
has shown that in median seeing conditions at Paranal, and with three
natural guide stars of magnitude 13 or brighter, it will be possible
to obtain a corrected field of view of $1'\times1'$ with an improved
and nearly constant PSF, and a Strehl ratio of a few \% at
0.6~\micron. This raises the exciting possibility to observe HDF-like
deep fields from the ground at a resolution similar to that of
HST.\looseness=-2

The number of actuators in deformable mirrors is currently limited to
a few hundred. This number will increase by almost an order of
magnitude in the next generation of deformable mirrors which use
micro-mirror arrays. This will considerably enhance the performance of
AO, in particular at shorter wavelengths.  In a further step, the
multi-layer oriented MCAO technique which will use more guide stars
and a larger field of view for the ground layer, should be able to
reach a Strehl ratio of 20\% at 0.5~\micron.  Example PSFs are shown
in Figure \ref{fig:stars_psf}.

\subsection{Advanced slicers}
\label{sec:slicers}

Three different concepts of IFS are in use: lenslet IFSs such as
\tiger, \oasis\ or \sauron\ (\cite{tiger,oasis,sauron}), lenslet+fiber
IFSs such as \vimos\ (\cite{vimos}) or \gmos\ (\cite{gmos}), and
slicer IFSs such as \threed\ (\cite{threed}) and \sinfoni\
(\cite{sinfoni}). Each concept has its pros and cons, but not all of
them can be expanded to a very large number of spatial elements and
not all have a high packing efficiency. The advanced slicer technique
pioneered by the Durham group (\cite{slicers}) is the most promising
in terms of packing efficiency (nearly 100\%) and overall size. It is
also one of the concepts selected as an alternative for a MEM design
for ESA's NIR spectrograph on NGST (\cite{ifmos}). Two prototypes are
being tested. Application of this concept to visible wavelengths is
being investigated at Lyon in collaboration with LAS and ESO.

\subsection{Tradeoffs}
\label{sec:tradeoff}

The main requirements for \mifs\ are summarized in Table
\ref{tab:Table1}: a wide field of view, high spatial resolution and
large simultaneous spectral range.  To meet these requirements, a very
large number of detector elements is needed, since in any IFS, the
total number of pixels is the number of spatial elements times the
number of spectral elements times the packing efficiency of the
IFS. The baseline already has 90,000 spatial elements. The number of
spectral elements is driven by the requirement of maximum simultaneous
spectral coverage (limited to an octave for a grating spectrograph)
and the necessary spectral resolution. The latter is set by the
science goals (\S\ref{sec:science}) but is also contrained by the need
to avoid the bright sky-emission lines in the red part of the
spectrum. To cover 0.6-1.0~\micron\ at a resolving power of 1500, one
needs 1600 pixels of 2.5 \AA. This translates to a total of 144
Megapixels for an instrument with 100\% packing efficiency.

\begin{table}[t]
\caption{Main requirements}
\begin{center}
\begin{tabular}{lll}
\hline
& \bf{Baseline} & \bf{Goal} \\
\hline
\noalign{\smallskip}
Field of View \& Sampling & 60\arcsec$\times$60\arcsec @ 0\farcs2 & 
120\arcsec$\times$120\arcsec @ 0\farcs2\\
& 3\arcsec$\times$3\arcsec @ 0\farcs01 & 
6\arcsec$\times$6\arcsec @ 0\farcs01\\
Wavelength range & 0.6-1.0~\micron & 0.35-1.0~\micron\\
Resolving power & 1500-3000 & 1500-5000 \\
Total throughput & 0.15 & 0.20 \\
\hline
\end{tabular}
\vskip -1.5\baselineskip
\end{center}
\label{tab:Table1}
\end{table}

Selection of the spectral range is a matter of science goals,
competitiveness, and technological limitations. Because of redshift,
the study of distant galaxies naturally benefits from the
infrared. However, an 8--10~m class ground-based telescope is not
competitive with NGST at $\lambda > $ 2.2~\micron\footnote{This is
true at low to medium spectral resolution. NGST can study only
relatively bright mid-IR sources at resolving powers R $>$
10000.}. The 1--2.2~\micron\ window might be a compromise: galaxies
can be studied to $z \leq 5$ using diagnostic lines such as \OII, and
MCAO will perform better in this wavelength range than at shorter
wavelengths.  However, the price per infrared detector pixel is
excessive, and their performance in readout noise and dark current,
despite their continuous improvement, is not as good as achieved by
CCD detectors. Furthermore, such a spectrograph would have to be
cooled at $\lambda >$ 1.6~\micron, which is difficult to achieve for
an instrument as large as the one envisioned here.

The 0.6--1~\micron\ wavelength range provides the best solution. MCAO
will give a significant improvement in spatial resolution, and the
technology to build and operate a large instrument is available.
Galaxies with $z \leq 7$ are accessible via the Ly$\alpha$ emission
lines, while medium to low-$z$ galaxies can be studied in detail using
the \OII\ or \Ha\ emission lines. At $\lambda < $ 0.6~\micron, MCAO
will have difficulties to produce increased spatial resolution, and
the high-$z$ galaxies cannot be observed. However, the \mifs\ concept,
even at the spatial resolution set by natural seeing, is still very
competitive and an extension of the 0.6--1.0~\micron\ baseline to
shorter wavelengths should be a design goal (Table~\ref{tab:Table1}).

In principle the diffraction limited core of the MCAO PSF should be
critically sampled, i.e., 0\farcs01 at 0.8~\micron. However, MCAO will
deliver at best a Strehl ratio of 10-20\%, and such fine sampling is
impractical given the size and faintness of the targets.  We therefore
selected a low-resolution spatial sampling of 0\farcs2. This is
compatible with the measured size of very faint distant galaxies
(typical half-light radius 0\farcs1). With a field of view of
$1'\times 1'$ sampled at 0\farcs2, the total number of detector pixels
is already very large. For individual bright targets we selected a
$3''\times3''$ field, sampled at 0\farcs01, to make optimal use of the
MCAO PSF.\looseness=-2

\begin{figure}[t]
\begin{center}
\includegraphics[width=.7\textwidth]{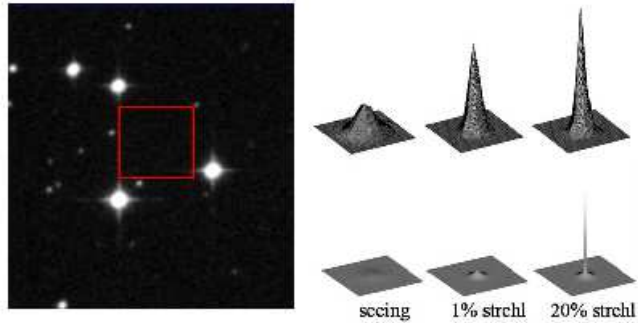}
\end{center}
\vskip -6truept
\caption[]{Left: An example of a possible \mifs\ 3D deep field
location (red box is $1'\times1'$).  Right: LR (upper box) and HR
(lower box) simulated \mifs\ PSF for median seeing (left), 1\%
(center) and 20\% (right) Strehl ratio.}
\label{fig:stars_psf}
\end{figure}

\subsection{Modular concept} 
\label{sec:concept}
The baseline instrument requires a 21k$\times$21k detector. This
cannot match one monolithic spectrograph, so we will split the
instrument into a number of modules. This also reduces the cost. The
number and size of the modules will be selected during phase A:
various choices are possible, from 17 modules with a 2k$\times$4k
detector each to a smaller number of modules with a larger detector.
Optical quality, volume, weight and cost will all be part of this
tradeoff.

\mifs\ will be set up at the VLT Nasmyth focus.  The MCAO module will
feed the $1'\times1'$ field of view into the preoptics. This enlarges
and splits the field of view into a number of separated beams. Two
enlargers define the two \mifs\ modes: low resolution with 0\farcs2
sampling, and high resolution with 0\farcs01 sampling.  Each sub-field
of view is then fed into an IFS module, which consists of an advanced
slicer, a classical grating spectrograph and a CCD detector.

\subsection{Performance}
\label{sec:performance}
We have computed the \mifs\ performance for extended high-$z$ objects
with 0\farcs1 half-light radius and an exponential surface brightness
distribution. The objects were convolved with the MCAO PSF provided by
ESO. We took a conservative approach, and assumed a Strehl ratio of
only 1.5\%. Even with this very modest Strehl ratio, a gain of 3.8 in
encircled energy within a 0\farcs2 pixel is obtained. We assumed a
total throughput of 0.15 and a detector readout noise and dark current
of respectively 3e$^{-}$ and 2e$^{-}$s$^{-1}$.  Integration time was
set to 80 hours split in 80 exposures of 1 hour. The typical sky
brightness at Paranal was used. The resulting limiting magnitude and
flux given in Table \ref{tab:Table2} are for a summation radius of
0\farcs3 and for a S/N of 5 per resolved spectral element (2
pixels). Table \ref{tab:Table2} also reports the limiting magnitudes
after summing over 10 pixels in the spectral direction which gives an
effective spectral resolution of 160.

Detailed analysis of the noise shows that \mifs\ is sky-photon-noise
limited. It is thus possible to coadd spatial and/or spectral pixels a
posteriori without loss of performance. However, to reach this
performance in practice, and sum exposures up to 80 hours of
integration time, requires a very stable instrument and full control
of all possible systematic effects. This would be very hard to do with
a multi-slit spectrograph but can be achieved more easily with an
IFS because of its complete spatial sampling of the field of view. For
example, it is possible to optimally coadd all exposures taking into
account the unavoidable differences in spatial resolution of the
individual exposures.

\begin{table}[t]
\caption{Estimated \mifs\ limiting magnitude for extended objects (0\farcs1)}
\begin{center}
\begin{tabular}{|l|rr|rr|}
\hline
\bf{Res.} & \multicolumn{2}{|c|}{\bf{R band}} & 
\multicolumn{2}{|c|}{\bf{I band}} \\
& R mag & F (\ergs) & I mag & F (\ergs) \\
\hline
1600 & 26.7 & $3. 10^{-19}$ & 25.2 & $5. 10^{-19}$ \\
160 & 28.2 & & 26.5 & \\
\hline
\end{tabular}
\vskip -1.5\baselineskip
\end{center}
\label{tab:Table2}
\end{table}

\section{Scientific objectives}
\label{sec:science}

Integral-field spectroscopy is the observing mode optimised for study
of resolved objects of every type. Thus, \mifs\ will be the instrument
of choice for studies ranging from Solar System objects, including
planetary weather, asteroid surfaces, and cometary activity, star
forming regions, including objects with proto-stellar and
proto-planetary disks, late stages of stellar evolution, especially
extreme AGB mass loss ($\eta$ Carina) and planetary nebulae, supernova
remnants, galactic nuclei, starbursts, AGN and merging galaxies, etc.
We restrict discussion to the study of the formation and evolution of
high-$z$ galaxies, the key driver for the \mifs\ low-resolution mode,
and to the unprecedented opportunities for studies of galactic nuclei
and young stellar objects provided by the high-resolution
mode. \looseness=-2

\subsection{High-redshift galaxies}
\label{sec:highz}
In the next five years, much progress will be made in our
understanding of the global properties of galaxies up to
$z$$\sim$1--2, through systematic studies with multi-object
spectrographs with high multiplex capabilities such as \vimos.
Resolved spectroscopy is required to measure the mass distribution and
stellar population mix in the galaxies.  Such measurements will remain
very difficult to make since these galaxies are both very faint {\em
and} very small. Even the new generation of multi-slit spectrographs
cannot afford to have 0\farcs2 slit width and thus will not, or just
barely, resolve the interesting galaxies.  For the $z$=3--4 range, and
presumably also for the $z=$0.8--1 range, \mifs\ will allow us to take
the next step, i.e., spectrally resolve the brighter galaxies. This is
the same step that has revealed so much about nearby galaxies in the
past 30 years, and allows addressing many fundamental properties
directly: (i) how are stellar populations distributed in $z$=0.8--1
galaxies (do we find young `kinematically distinct cores'?)~(ii)
detection of stellar absorption lines in $z$=3--4 galaxies, and their
kinematics.

\begin{figure}[t]
\begin{center}
%\vskip 6.3truecm
\includegraphics[width=.7\textwidth]{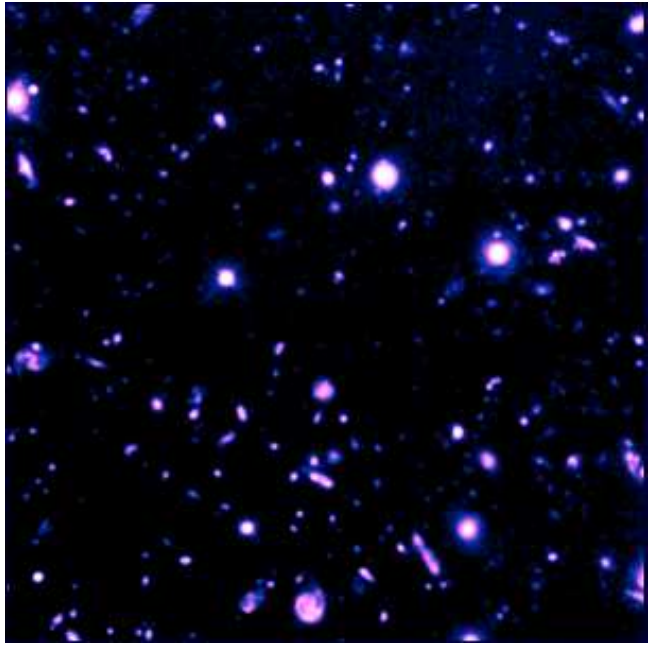}
\end{center}
\caption[]{Simulated HDF-\mifs\ 3D deep field. Only Lyman break
galaxies are shown here.}
\label{fig:hdf}
\end{figure}

\mifs\ with its IFS capability will be able to perform such a study in
its low-spatial resolution mode, and reach galaxies down to R=26.7
(28.2 at lower spectral resolution).  With this limiting magnitude,
\mifs\ will be able to study 76\% of the $z<1$ galaxies, 67\% of the
$1\!<\!z\!<\!3$ galaxies and 41\% of the $z\!>\!3$ galaxies in the HDF
(based on the photometric redshifts in the HDF-North).  We estimate
that 80\% of the detected objects with $1\!<\!z\!<\!3$ will have at
least a few resolved spectral elements. The number of resolved
elements is likely to be larger in the case of detected emission
lines.  Figure~\ref{fig:hdf} shows a simulated \mifs\ deep field. It
was computed from a $1'\times1'$ window of the HDF-North, convolved
with the appropriate MCAO PSF (\S\ref{sec:performance}) and binned
with a 0\farcs2 pixel.  \looseness=-2

\subsection{Search for Ly$\alpha$ emitters}
\label{sec:lya}
The \mifs\ spectral window allows a blind search for Ly$\alpha$
galaxies in the range $4\!<\!z\!<\!7$. With a detection limit of
3--5$\times$10$^{-19}$\ergs, \mifs\ will explore a completely new
parameter regime, and will reveal the emission-line objects that are
too faint to be detected with broad-band imaging: an object with
E(Ly$\alpha$) of 3$\times$10$^{-19}$ \ergs\ and an equivalent width of
100 \AA\ would have $R\sim 29.7$.\looseness=-2

The expected number of objects that \mifs\ will detect in a field of
$1'\times 1'$ is difficult to estimate, as it requires extrapolation
of the current narrow-band imaging detections at
2$\times$10$^{-17}$\ergs\ (\cite{hu}) by two orders of magnitude. At
the present detection level, Ly$\alpha$ galaxies are relatively rare:
1--3 objects per arcmin$^2$ only, but narrow-band imaging surveys
typically probe only a range of 0.05 in $z$, while \mifs\ covers the
range $4\!<\!z\!<\!7$ in a single exposure.

In order to obtain a more reliable estimate, we used a hybrid model of
hierarchical galaxy formation (\cite{hatton}), which follows the
history of dark matter in the standard $\Lambda$ cold dark matter
scenario using large N-body simulations. The history of baryons is
computed with semi-analytic recipes which include cooling, star
formation, merging, chemical evolution and dust absorption.  We take
the specific dust absorption on Ly$\alpha$ into account empirically,
by rescaling the model predictions to fit the observed numbers of
Ly$\alpha$ emitters at $z\!\sim\!3.4$ (\cite{hu}).  This leads to dust
absorption by a factor of eight, with a nearly negligible resonant
scattering. The simulations have a mass resolution of
$10^{10}$~\Msun. The predictions of the model reproduce the observed
properties of $R<26$ galaxies in the HDF, and are limited to
E(Ly$\alpha$) $>$ $10^{-18}$~\ergs.  At this level, the model already
predicts about 150 Ly$\alpha$ emitters per arcmin$^2$ in the range
$4\!<\!z\!<\!7$ (and an additional 675 for $2\!<\!z\!<\!4$).  90\% of
the 152 high-redshift objects are too faint to be detected in
broad-band imaging.  While one should be cautious not to overestimate
the accuracy of the model predictions, there are two reasons to be
optimistic: (i) the detection limit of \mifs\ will be a factor three 
better than used in the simulations, and (ii) some high-redshift
Ly$\alpha$ galaxies were already discovered serendipitously in
long-slit HDF follow-up observations: Dawson et al. (\cite{dawson})
discovered 11 galaxies at $3\!<\!z\!<\!6$ of which only one does not
have detectable Ly$\alpha$ emission.  Five of these Ly$\alpha$
emitters have no detectable continuum (I$_{AB} >$ 25). This is very
encouraging since a blind search with a long slit is particularly
inefficient.\looseness=-2

We conclude that \mifs\ will be able to observe the population of
$z>4$ galaxies. Most of the objects will be {\em new}, and cannot be
observed with present instrumentation. This population of new galaxies
should be different from the population of Lyman break galaxies: they
are low-mass galaxies experiencing strong star formation, and are the
progenitors of present day normal galaxies.

\subsection{Galactic nuclei}
\label{sec:sbh}
HST has revealed that the nuclei of many nearby normal galaxies
harbour a supermassive black hole, as well as cusped density profiles,
stellar and/or gaseous disks. The black hole causes a strong gradient
in the mean velocities of stars and gas, and a central increase of the
velocity dispersions, which can be modeled to measure the black hole
mass (e.g., \cite{g00,vdM98}). The nuclear properties turn out to
correlate with the global structure of the galaxies (e.g.,
\cite{f97,fm00,g00}), so that studying the nuclei provides important
constraints on galaxy evolution.

In the center of M31, the distribution of light and the stellar
kinematics as seen by \oasis\ and \stis\ is complex (\cite{bac01}),
and similar asymmetries no doubt occur in the nuclei of other
galaxies.  At intermediate and larger scales, the ongoing \sauron\
survey of the two-dimensional kinematics and stellar populations of
nearby galaxies already shows that departures from axisymmetry are
likely to be the rule rather than the exception (\cite{dz01}).
Unfortunately, high resolution measurements of the two-dimensional
stellar kinematics is very demanding for HST given its limited
aperture. In order to make progress, one needs an IFS with much higher
spatial resolution than is available today. \sinfoni, which will be on
the VLT in 2004, will deliver 0\farcs06 spatial resolution at
2~\micron\ over a $2''\times2''$ field.  The next step would be
achieved by \mifs\ in its high resolution mode (0\farcs01) given that
MCAO achieves a Strehl ratio of 20\% or higher (so that the
diffraction peak can be differentiated from the outer halo). The field
of view of $3''\times3''$ is well-matched to the size of the nuclear
stellar and gaseous disks, and the decoupled kinematic cores seen in
many nearby objects.

\subsection{Young stellar objects}
\label{sec:yso}

Many young solar mass stars are surrounded by Keplerian disks with
masses and sizes similar to our primitive Solar System
(\cite{Beck}). The evolution of physical conditions in these disks
eventually governs planet formation in our Galaxy. A crucial unknown
is the launch point of the collimated winds observed in these
systems (somewhere in the disk, or the stellar magnetosphere?),
and the resulting impact on the internal structure and evolution of
the disk. Spectro-imaging in atomic forbidden lines near 0.6$\mu$m
with \tiger, \oasis\ and \stis\ revealed the two-dimensional wind
kinematics and excitation conditions down to 0\farcs1 (15~AU) of the
star (\cite{Bac00,L97}). A resolution of $\simeq 0.01''$ is
needed to resolve the regions within 1~AU where wind acceleration and
collimation take place, and to constrain the wind origin.  Since atomic
winds emit very little in the near-infrared, neither \sinfoni\ nor
NGST are likely to bring significant advance. The high-resolution mode
of \mifs\ will allow this decisive step, in a field of view ideally
matched to the bright inner wind regions. A resolving power of 5000
would be crucial for resolving velocity gradients in the wind
acceleration region. In its low-resolution mode, \mifs\ will probe
in unprecedented detail the two-dimensional kinematics and excitation
conditions of large shock fronts and Herbig--Haro objects at larger
distance from the star, providing key constraints on the wind magnetic
field.\looseness=-2

\section{Phased development and risks}
\label{sec:risk}
Given its key importance for the construction of extremely large
telescopes, MCAO is a strategic area of research and development.  ESO
is currently building a demonstrator that should be on sky in
2003. This is essential to acquire expertise in MCAO but the
demonstrator will not have science capabilities. 

The next step will be to build a fully operational MCAO system using
three wave-front sensors on natural guide stars and two low-order
deformable mirrors. This should easily achieve the Strehl ratio of
1.5\% at 0.8~\micron\ assumed in \S\ref{sec:performance}. When combined
with \mifs\ in its low-resolution mode, this will provide the deep
fields needed for the programs described in \S\S\ref{sec:highz} \&
\ref{sec:lya}. We have found many areas of blank sky that fulfill the
requirements of having guide stars brighter than $R=13$ with the
correct geometrical configuration, and at high galactic latitude. An
example located near the HDF south is shown in Figure \ref{fig:stars_psf}.

The \mifs\ high-resolution mode demands higher performance from
(MC)AO, and the science goals described in \S\S\ref{sec:sbh} \&
\ref{sec:yso} will only be achieved with a second generation AO
system, so will presumably come a few years later.

In case MCAO is late achieving the first phase goals, \mifs\ can be
used in natural seeing. Even in this configuration it would be a
unique instrument in terms of field of view, spatial sampling and
spectral coverage and resolution. With 0\farcs7 median seeing, there
is no hope to resolve the tiny high-redshift galaxies, but \mifs\ will
be able to obtain significant results in integrated spectroscopy. Its
performance will be less than those given for phase 1, but, e.g., the
search for Ly$\alpha$ emitters would still be unchallenged by other
instruments.

In terms of the design of the IFS itself, most of the required
technology is in hand. The feasibility of advanced slicers was
demonstrated by the ongoing prototype studies. Instrument stability
will be eased given the small number of moving parts. The cost can be
kept within reasonable limits by working with optical industry to
design an IFS module that would minimize the cost of the
optics.\looseness=-2

\section{Conclusions}
\label{sec:conclusions}

\mifs\ is a true {\em second generation} IFS. The first generation is
being used to study in great detail individual objects that were
previously discovered with imaging surveys. With its large field of
view and simultaneous spectral coverage, \mifs\ combines the discovery
potential of an imaging device with the measuring capabilities of a
spectrograph, while taking advantage of the increased spatial
resolution provided by MCAO. This makes it a unique tool for
discovering objects that cannot be found in imaging surveys.  The
phased development described in \S\ref{sec:risk} minimizes risk, and
guarantees that the main scientific goals will be
achieved.\looseness=-2

Providing new contraints on the formation and evolution of galaxies at
high redshift is the primary scientific objective we have used to set
the specifications of the low-resolution mode of \mifs. The
spectroscopic deep fields obtained with \mifs\ will constitute a
tremendous treasure of information which will become a lasting
reference, similar to the Hubble Deep Fields of today.  \mifs\ deep
fields will cover the visible spectral window and will perfectly
complement the NGST (IR) and ALMA (mm) spectral windows.
Multi-wavelength coverage of the same fields by these three facilities
will provide nearly all the measurements we will need in order to
answer the critical question of the formation of galaxies.

Using the best spatial resolution that can be achieved at one VLT unit
to look at the environment of, e.g., supermassive black holes and
young stellar objects is the main driver for the high-resolution mode
of \mifs. But these two subjects are only a tiny fraction of the
science that could be carried out with such instrumental
capabilities. Any object that has spectral features in the visible
range and needs high spatial resolution is a potential target.

The first generation VLT instruments provide nearly complete coverage
of observational parameter space (spatial/spectral resolution and
wavelength range). \mifs\ expands this to high spatial resolution
spectroscopy at visible wavelengths.  It has a large potential for
discoveries, builds upon the leading role that Europe has in
integral-field spectroscopy, maximizes the return from the new
developments in adaptive optics, and will keep the VLT competitive for
the next decade.\looseness=-2

In the longer term, ESO plans to construct OWL. MCAO is on the
critical path. Achieving regular scientific use of MCAO with a VLT
instrument would be a key step towards realising OWL. OWL MCAO is not
only challenging by itself, but also for the required instrumentation.
Much of the science case for OWL requires high spatial resolution
spectroscopy. As 1 arcsec$^2$ requires 630,000 spatial elements to
properly sample the OWL PSF, the constraints on the spectrographs will
be enormous. While still short of achieving the requirements for OWL,
\mifs\ with its 90,000 spatial elements is a major step beyond the
capabilities of the present IFS: \mifs\ has 30 times the number of
spatial and spectral elements of \vimos, and 180 times those of
\sinfoni.

\medskip
\noindent
{\bf Acknowledgments.} We thank Norbert Hubin, Rodophe Conan and the
ESO AO division for providing very valuable MCAO performance
simulations. This work was supported by the Programme National
Galaxies from INSU/CNRS.

%INDEX%%%%%%%%%%%%%%%%%%%%%%%%%%%%%%%%%%%%%%%%%%%%%%%%%%%%%%%%%%%%%%%
% Please check with the editor of your book whether he plans to
% include a "mutual" subject index - if so, please code your entries
% in the standard syntax. For your own purposes you may print your
% "personal" index by using the following commands:
%
%\clearpage
%\addcontentsline{toc}{section}{Index}
%\flushbottom
%\printindex
%%%%%%%%%%%%%%%%%%%%%%%%%%%%%%%%%%%%%%%%%%%%%%%%%%%%%%%%%%%%%%%%%%%%%

\end{document}